\documentstyle[11pt]{article}

\def\llt{{\hbox{$\scriptscriptstyle <$}}}
\def\lgt{{\hbox{$\scriptscriptstyle >$}}}
\def\sTME{{\hbox{$\scriptscriptstyle {\rm TME}$}}}
\def\sTMG{{\hbox{$\scriptscriptstyle {\rm TMG}$}}}
\def\sCS{{\hbox{$\scriptscriptstyle {\rm CSE}$}}}
\def\sCSG{{\hbox{$\scriptscriptstyle {\rm CSG}$}}}

\def\sEG{{\hbox{$\scriptscriptstyle {\rm EG}$}}}

\def\th{\theta} 
\def\hth{\hat{\mu}}
\def\pa{\partial}
\def\k{\kappa}
\def\g{\gamma} \def\G{\Gamma}
\def\a{\alpha}
\def\b{\beta}
\def\d{\delta} \def\D{\Delta}
\def\e{\epsilon}

\def\k{\kappa}
 
\def\m{\mu}
\def\n{\nu}

\def\p{\pi}

\def\ra{\rightarrow}

\renewcommand{\baselinestretch}{1}
\begin{document}

\noindent\hspace*{5in}BRX TH--337\\
\hspace*{5in}DAMTP R93/5\\
\hspace*{5in}ADP-93-204/M18

\vspace*{-.2in}

\begin{center}
{\bf ULTRA--PLANCK SCATTERING IN {\boldmath $D$}=3 GRAVITY
THEORIES}

\vspace{.2in}
S. Deser \\
{\it Department of Physics}\\
{\it Brandeis University} \\
{\it Waltham, MA 02254-9110, USA}

\vspace{.1in}

J. McCarthy\\
{\it Department of Physics}\\
{\it University of Adelaide}\\
{\it Adelaide, SA 5001, Australia}

\vspace{.1in}

Alan R. Steif\\
{\it DAMTP}\\
{\it Cambridge University}\\
{\it Silver St.}\\
{\it Cambridge CB3 9EW, U.K.}

\vspace{.3in}

ABSTRACT
\end{center}

\renewcommand{\baselinestretch}{2}
\small
\normalsize
We obtain the high energy, small angle, 2-particle gravitational
scattering amplitudes
in topologically massive gravity
(TMG) and its two non-dynamical constituents, Einstein and
Chern--Simons
gravity.  We use 't Hooft's approach, formally equivalent to
a leading order eikonal  approximation:
one of the particles is taken to scatter through the classical
spacetime
generated by the other, which is idealized to be lightlike.  The
required geometries are derived in all three models;
in particular, we thereby provide the first explicit asymptotically flat
solution generated by a
localized source in TMG.  In contrast to $D$=4, the metrics
are not uniquely specified, at least by naive asymptotic requirements -- an
indeterminacy mirrored in the scattering amplitudes.  The eikonal
approach does provide a unique choice, however.
We also discuss the discontinuities
that arise upon taking the limits, at the level of the solutions, from
TMG to its constituents, and compare with the analogous topologically
massive vector gauge field models.

\newpage

\noindent{\bf Introduction}

The scattering of two gravitationally interacting Planck energy particles
has been studied in recent years from several, quite
different, points of view \cite{ACV1,tH2,VV3}.
To leading order in the large $s$, small  $t$  expansion these
calculations agree, and also coincide with those of the usual
eikonal approximation obtained by summing a ``leading'' subset of
Feynman
diagrams (see, {\it e.g.}, \cite{KO14} and references therein).
The essence of the approximation is graphically expressed
in 't Hooft's approach  \cite{tH2}: choose a
frame in which one of the particles is essentially light-like and so
generates an impulsive plane-fronted shock wave spacetime through
which the other scatters as a (quantum mechanical) test body.  It
is quite straightforward to apply this method to other models, at least
in principle:
first calculate the
spacetime
generated by the rapid mover and then determine the evolution of the
initial free state of the other particle in this (background) geometry.
The scattering amplitude is of course just the overlap of the outgoing
state with a given final-state free wave function.
Because the geometry is impulsive, its field is ``piece-wise''
pure gauge -- or equivalently a pure gauge
but singular metric -- and the scattering is essentially of
Aharonov--Bohm
type.  More recently, the Verlindes \cite{VV3} suggested a
Lagrangian-based derivation of this approximation in $D$=4 gravity,
using a scaling argument that reduces the strong-coupling sector to
a topological field theory in which the above semi-classical
dynamics is the lowest order effect.  Although it is not clear how these
arguments can be extended beyond the lowest order, they do
show heuristically, at least, how to freeze out the gravitational quantum
modes at the level of the action.  Their ideas have been applied to
$D$=3 Einstein gravity \cite{Zeni}, and compared \cite{KO36} with
other approaches there.  In all cases, the question of
whether one thereby obtains the truly {\it dominant}
contributions is still
under discussion \cite{ACV1,KO25}.

It is clearly of interest, in order to further assess their validity, to test
these ideas on other gravitational
systems, including those in which there are no gravitons to be frozen
out in the first place, but that are limiting cases of a dynamical theory.
In this respect, $D$=3 models are especially useful, comprising as they
do both of the nondynamical examples -- Einstein gravity
and conformally invariant pure Chern--Simons gravity (CSG) --
in which there are no gravitons at all \cite{DJT6},
as well as their sum, topologically massive gravity (TMG), which does
have
a (single massive helicity 2) dynamical excitation \cite{DJT6}.
For each theory, we will first obtain the
geometries generated by a null source
(which we will call a photon for brevity)
by showing that, just as in $D$=4, they are of shock wave form.  In
particular, we will obtain
the metric for the full
nonlinear TMG field equations.  This is of interest in its own right, as
it
provides the first known asymptotically flat solution corresponding to
a localized
source in TMG; even the ``Schwarzschild'' solution is only known to
linearized
order there \cite{who7}.  [In \cite{DS8} it was shown that this complete
plane
wave geometry may in fact be obtained by an infinite
boost from that linearized metric.]
We will then exhibit the required scattering amplitudes and show
that they contain intrinsic ambiguities peculiar to the global aspects of
$D$=3
gravities; we propose a (heuristic) choice based on the eikonal
prescription.  It also turns out that the limits of
TMG that yield its constituent actions can lead to singularities.  We will
analyze this question and compare with the analogous topologically
massive vector gauge
models.

\noindent{\bf Plane Wave Geometries}

For definiteness, take the photon source to be right-moving along
$x$, with energy $E$.
The distinctive property of its energy-momentum tensor (in any $D$)
is that it has only one non-vanishing component,
$T_{uu} =  E\delta (\mbox{\boldmath $y$})\delta (u)$, in the usual
lightcone and
transverse coordinates
$(u,v,\mbox{\boldmath $y$}) \equiv (t - x, \; t + x,\mbox{\boldmath
$y$})$.
This motivates the following
ansatz for the metric in harmonic gauge,
\begin{equation}
ds^{2} = ds^2_0 + F(u,\mbox{\boldmath $y$}) du^2 \; , \hspace{.4in}
ds^2_0 = -dudv
+ d\mbox{\boldmath $y$}^{2} \; ,
\label{L1}
\end{equation}
and indeed the Einstein tensor  of (\ref{L1}) reduces to the simple
linear form
\begin{equation}
G_{\mu\nu} = - \textstyle{\frac{1}{2}} \nabla^{2}_T F \, l_\m l_\n \,
,
\hspace{.4in}
l_\m \equiv \pa_\m u \, ,
\label{L2}
\end{equation}
in terms of the transverse $D-$2 flat-space Laplacian.
[Our conventions are: $R_{\mu\nu} \sim + \pa_\a \G^\a_{\mu\nu}
\; , \;\; \e^{txy} = +1$.]
Henceforth we specialize primarily to $D$=3, where we simply
denote
$y$-differentiation by a prime.  Clearly, only the component $G_{uu}$
(or equivalently only $R_{uyuy}$, since
Riemann and Einstein tensors are mutual double duals in $D$=3) fails
to vanish.  In addition, the
more complicated third-derivative (traceless, conserved, and symmetric)
Cotton--Weyl tensor,
$C^{\mu\nu} \equiv (-g)^{-1/2}
  \e^{\m\a\b} \, D_{\a}
(R^\n_\b - \textstyle{\frac{1}{4}} \, \d^\n_\b \, R) $,
also
simplifies nicely under this ansatz, becoming
\begin{equation}
C_{\mu\nu} = \textstyle{\frac{1}{2}} F^{\prime\prime\prime} \; l_\m
l_\n \; .
\label{L3}
\end{equation}
Consequently, the (parity violating) field equations of TMG,
\begin{equation}
G_{\mu\nu} + \textstyle{\frac{1}{\mu}}
\, C_{\mu\nu} =
-\kappa^{2} \, T_{\mu\nu} \, ,
\label{L4}
\end{equation}
reduce to a single third-order linear equation for the metric $F$:
\begin{equation}
F^{\prime\prime} - \m^{-1}\,F^{\prime\prime\prime}  =
2 \kappa^2 E \, \d (y) \d (u) \; .
\label{L5}
\end{equation}
Here $\kappa^2$ is the Einstein constant (with dimensions of inverse
mass  in
$D$=3) and $| \mu |$ is the mass of the TMG excitation (our convention
is $\m > 0$).
[Note the parity violation of TMG, as reflected in the appearance
of the orientation $\epsilon_{\m\n\rho} $ in the equations of motion;
changing the sign of $\m$ is equivalent to performing a parity
transformation.]
  We recall for future
reference that the sign of
$\k^2$ here is necessarily opposite to that usually taken in Einstein
theory in order that the TMG excitations be non-ghost \cite{DJT6}.

The other two gravity models discussed in the
introduction are
contained as limits of TMG,
whose field equations (\ref{L4}) reduce to the
(ghost sign) Einstein equations as $\m \rightarrow \infty$, while
those of pure Chern--Simons
gravity $(C_{\mu\nu} = - \m \k^2 T_{\mu\nu})$ are
reached as $ \mu\ra 0,$ $\kappa^2\rightarrow \infty$, keeping
$\tilde{\mu} \equiv \m\k^2$ fixed.  We will see, however, that taking
these limits in the explicit solutions of TMG can be more delicate.

The curvature is easily obtained by one integration of (\ref{L5}),
\begin{equation}
G_{uu} = -\textstyle{\frac{1}{2}}\, F^{\prime\prime} = -\m \kappa^{2}
E
e^{\m y} \th(-y) \d (u)  \; ,
\label{L6}
\end{equation}
upon choosing the homogeneous solution for $F^{\prime\prime}$
so as to obtain asymptotic flatness in $y$, {\it i.e.}, excluding any
source-free graviton part of the curvature.
Integrating (\ref{L6}) then yields the metric $F$,  which we
write as
\begin{equation}
F = 2E\D_T (y)\d (u) \; , \label{L7}
\end{equation}
in terms of a Green's function $\D_T$ for the transverse kinetic
operator of (\ref{L5}),
\begin{equation}
\D_T (y) = \k^2 (D^2 -\m^{-1}D^3 )^{-1} \d (y)
= \frac{\k^2}{\m} \, [\frac{1}{\m -D} + \frac{\m}{D^2} + \frac{1}{D} ]
\d (y) \; , \;\;\;\;\;\;\;\;  D \equiv d/dy \; .
\label{L7a}
\end{equation}
Having specified (\ref{L6}) by imposing asymptotic flatness, the most
general homogeneous
solution $H$ that we may add to $\D_T$ is that  of
$H^{\prime\prime} = 0$, namely
$H = Ay + B$.
In the following we shall take
\begin{equation}
\D_T(y) = \D_{T}^\sTMG(y) =
\k^2 [\th (-y) \m^{-1}  e^{\m y}  + \epsilon (y)
\textstyle{\frac{1}{2}} (y +\m^{-1})] \; ,
\label{L9}
\end{equation}
but we will return to the significance of this particular choice, and to
the role of the $H$ ambiguity.
The geometry given by (\ref{L7}) and (\ref{L9}) is the advertised
solution of the full
nonlinear TMG field equations due to a localized (photon) source; as we
saw, it is
correspondingly flat
(though the metric is not manifestly cartesian) at infinity and
nonsingular away from the photon.

Let us consider the properties of the curvature (\ref{L6}) and the
relevant
part, (\ref{L7}), of the metric in our various models,  with the given
choice of homogeneous solutions.  For TMG itself, the curvature fails to
vanish only on the  $y < 0$ half of the $u=0$ null plane, while the
metric
has been permitted to be non-cartesian also at $y > 0$.
In Einstein theory, which is (\ref{L5}) at $\m = \infty$, flatness
reigns everywhere outside the source.  If we take the $\m \rightarrow
\infty$ limit of (\ref{L6}), it agrees since
$\m e^{\m y} \th (-y) \rightarrow \d (y)$
(for smooth enough test functions); likewise (\ref{L9}) limits to
the $y$-symmetric form
$\D_T = \k^2  |y| /2$.  Although
$F$ is neither in pure gauge nor in asymptotically Cartesian form, we
shall see that it can be set to zero locally by (singular) coordinates
choices; globally a conical space structure is unavoidably present and
will be quite relevant to our scattering problem.

The other limit discussed
above gives pure Chern-Simons gravity (CSG), to which only
a traceless $T^{\mu\nu}$ -- such as that of our photon --
may couple, since $C^{\mu\nu}$ is identically
traceless.  Also, since
$C^{\mu\nu}$ is in fact the Weyl tensor in $D$=3, the CSG metric is
only determined
up to a conformal factor and spacetime is now only conformally flat
outside the sources.  The field equation now reduces to
\begin{equation}
F^{\prime\prime\prime} = -2 \tilde{\m} E \d (y)\d (u) \, ,
\label{LI10}
\end{equation}
which integrates to
\begin{equation}
F^{\prime\prime} = - \tilde{\m} E \e (y) \d (u) \; ,
\;\; F = - {\textstyle\frac{1}{2}} \tilde{\m} E y^2 \e (y) \d (u)
\label{LI11}
\end{equation}
up to a constant in $F^{\prime\prime}$ and a homogeneous
$Cy^2 + Ay + B$ solution (here $H^{\prime\prime\prime} = 0$) in
$F$.  Note that no choice of $H$ can make $F^{\prime\prime}$
asymptotically flat (let alone flat for $y \neq 0$) in $y$; we have taken
the ``most symmetric'' option in (\ref{LI11}), but $F$ cannot in any case be
turned into a pure gauge.
Note also that the indeterminacy of the curvature is just the conformal
factor ambiguity; the resulting $Cy^2$ term in $F$
is an ambiguity over and above the effect of the homogeneous
solution $Ay + B$ common to all three models.
If we take the
CSG limits of (\ref{L6}) and (\ref{L9}) in TMG, we find that they diverge,
differing from (\ref{LI11}) by homogeneous solutions to be sure, but with
infinite coefficients.  Quite clearly this could have been avoided by
choosing for (\ref{L9}) the form
$\D_T = \th (-y) \k^2\m^{-1} (e^{\m y} - 1 - \m y)$, {\it i.e.},
by adding $H = - \frac{1}{2} \k^2 (y + \m^{-1})$ to (\ref{L9})
(incidentally, the Einstein limit would then still be finite and in
fact would be cartesian at positive $y$; in particular, it
is possible to choose a solution with finite limit in both directions!).
It may seem perverse to make the choice
(\ref{L9}) rather than the non-singular one; we have done so because
it corresponds to using the Feynman
propagator for the exchanged ``graviton'' in the eikonal perturbation
theory calculation.
[The boundary condition in
$\Delta_T$ is determined by the ``$i\epsilon$'' prescription chosen for
the
covariant propagator $D(x)$ integrated against the photon stress tensor.
Thus, in Einstein gravity, use of $D_{ret}$  leads to
$\Delta_T \sim y\theta (-y)/2$ instead of the $D_F$ value $y \epsilon (y)/2
= |y|/2$.]
We will return to this point after demonstrating the physical
consequences of
the ambiguity.  Suffice it to say here that (\ref{L9}) represents a
uniform
choice for all three models treated separately; this will make any
H-dependence in physical quantities even more striking.

\noindent{\bf Scattering}

Just like its $D$=4 Einstein counterpart \cite{AS9},
the TMG solution
(\ref{L9}) takes the form of an impulsive  plane-fronted wave, so
we can apply the analysis of {\cite{tH2}} to
study the small angle scattering of a particle in the background
generated by the photon.  To make this shock-wave character more
explicit, we first perform the
coordinate transformation
$v\rightarrow v + 2 E \D_T (y)\theta(u)$, after which the interval
takes the form
\begin{equation}
ds^2 = ds^2_0  -2 E \: \th (u) \D^\prime_T (y) dy \, du \,
{}.
\label{L10}
\end{equation}
That the metric is Minkowski for $u\neq 0$ is already manifest
for $u < 0$ in the
coordinates of (\ref{L10}); for $u > 0$ we write (\ref{L10}) as
\begin{eqnarray}
ds^{2} & = & - du \; d ( v + 2 \, E \, \D_T (y)) + dy^{2}
\nonumber
\\
& \equiv & -du \; dv_\lgt + dy^{2} \, .
\label{L11}
\end{eqnarray}
Thus the effect of the impulse is entirely summarized in the relation
between
the $x^\m_\llt$ and $x^\m_\lgt$ coordinates on the null $u-$plane,
namely by
\begin{equation}
u_\llt = 0 = u_\lgt \; , \;\;\; y_\llt = y_\lgt \; , \;\;\; v_\llt = v_\lgt
- 2 E\, \D_T (y_\lgt) \; .
\label{L12}
\end{equation}

Before the impulse, then, the test particle sees no field and the incident
wavefunction can be taken to be a plane wave with (on-shell)
momentum $p_\m$,
\begin{equation}
\psi_\llt = \frac{1}{(2\pi)^{3/2}} e^{i \; p \cdot x_\llt} \, .
\label{L13}
\end{equation}
After the impulse we have simply (still at $u = 0$)
\begin{equation}
\psi_\lgt = \frac{1}{(2\pi)^{3/2}} \exp  i[p_v (v - 2 E\D_T (y)) +  p_y y]
\; .
\label{L14}
\end{equation}
Decomposing (\ref{L14}) into plane waves in the out-region, we find for
the
scattering amplitude, \linebreak
$<k, {\rm out}|p , {\rm in}> = \d (p_v -k_v) T (q_y =p_y -k_y)$,
the expression
\begin{equation}
T(q_y)= \int {dy\over {(2\p)^2}}\exp i[q_y y - 2 p_v \, E \D_T (y)] \;
{}.
\label{L14a}
\end{equation}
For the kinematics discussed here, the Mandelstam variables are
$s = 4 | p_v|E $ and $t = - q_y^2$ since the momentum transfer is $q_y =
p_y - k_y $.
Thus the scattering amplitude is the usual
``leading'' eikonal expression \cite{KO14,KO25},
{\it i.e.}, the transverse Fourier transform of the exponentiated
``Coulomb'' potential.  The  amplitude (\ref{L14a}) can also be
understood in terms of an equivalent completeness argument.  Specifically,
let $G(x_1; x_2)$ be the propagator in flat spacetime ($x_1^{\m}$ and
$x_2^{\m}$ are spacetime $3-$vectors). By completeness,
the propagator from ($v_1$, $y_1$) on some initial  $u<0$ slice to
($v_2,$ $y_2$) on some  final $u>0$ slice is given by
\begin{eqnarray}
G(x_1; x_2) &=& \int_{u=0}
dvdy \, G(x_1 ; u_\llt, v_\llt , y_\llt ) G( u_\lgt, v_\lgt , y_\lgt ; x_2)
 \; ,
\label{LI18}
\end{eqnarray}
where the integral over $v$ and
$y$ is on the intermediate $u_\llt = u_\lgt = u = 0$ slice.
 Taking the Fourier
transform with incoming and outgoing  momenta $p_{\m}$ and
$k_{\m},$ yields
\begin{eqnarray}
G(p,k) & = &{1\over {(2\p)^3}}  \int dx_1 \, e^{ip \cdot x_1}
 \int dx_2 \, e^{ - ik \cdot x_2}
   \int_{u=0}
dvdy G(x_1 ; x_\llt ) G( x_\lgt ; x_2 )  \nonumber\\
&=&  G(p) G(k)
\int  dvdy \, e^{(ip \cdot x_\llt -  ik \cdot x_\lgt)}\; ,
\label{LI19}
\end{eqnarray}
Amputating the external propagators $G(p)G(k)$ then gives
(by the reduction formula) the scattering amplitude.  The result is
\begin{eqnarray}
<k, {\rm out}|p, {\rm in}>
& = & {1\over {(2\p)^3}}
 \int dvdy \exp { ( ip_v (v - 2  E \D_T(y)) + i p_yy - ik_v v - ik_y y
)}
\nonumber \\
& = & \d(p_v-k_v) \int {dy\over (2\pi)^2} e^{(i q_y y -2 i E p_v
\D_T(y))} \; ,
\label{LI20}
\end{eqnarray}
thereby  reproducing (\ref{L14a}).

The (TMG) Green's function $\Delta^{\sTMG}_T$ is, as we have seen,
only determined
up to
homogeneous solutions $H = Ay + B$.  Examining the
amplitude
(\ref{L14a}), it
is clear that for real $s$ the ambiguity generated by $B$ simply
amounts to an irrelevant constant
phase in the amplitude.  One also sees that adding $Ay$ to the Green's
function is equivalent to transforming the incoming momentum, $p_{\m}$, via
$p_y \ra p_y - \tilde A p_v,$ $p_v \ra p_v $ where $\tilde A \equiv
2EA$.  This
is a Lorentz
transformation
provided $p_u$ is transformed as well, according to
\begin{equation}
p_u ={p_y^2\over 4 p_v} \ra  \: {(p_y - \tilde{A} p_v)^2\over 4 p_v}
= p_u -
\textstyle{\frac{1}{2}}\tilde A p_y + \textstyle{\frac{1}{4}}\tilde A^2
p_v \;\; .
\label{LI21}
\end{equation}
The photon's momentum is obviously left invariant under
this transformation. Hence, the ambiguity associated with $A$
corresponds
to applying a Lorentz transformation in the entire in-region
({\it i.e.,} on both incoming particles).
[This result can also be seen directly from the original metric (\ref{L7})
and (\ref{L9}), where
adding  a linear term to the Green's function is equivalent to the coordinate
transformation
\begin{equation}
u \rightarrow  u \; , \;\;\;\;\;\;\;
v \rightarrow v+ \theta (-u) ( \tilde A y +
\textstyle{\frac{1}{4}}\tilde A^2  u + B ) \; ,
\;\;\;\;\;\;\;
y \rightarrow y+ \textstyle{\frac{1}{2}}\tilde A  u \theta (-u) \, ,
\label{L14b}
\end{equation}
 corresponding   to a Lorentz  transformation in  the $u<0$
halfspace.]
Clearly, performing  a Lorentz transformation  only in the in-region
does  {\it not} in general
leave the
S-matrix invariant.
Therefore, different choices of $A$ yield
different scattering amplitudes.  This indeterminacy is special to
$D$=3, in contrast to
$D$=4, where a unique choice is picked out
by demanding that the metric be asymptotically Cartesian.
There, in the impulsive
plane wave
metric \cite{AS9}, $\D_T (\mbox{\boldmath $y$})$
is the two-dimensional transverse Coulomb Green's function:
\begin{equation}
\D_T(\mbox{\boldmath $y$}) \sim
 \log \, \mbox{\boldmath $y$}^2   + H(\mbox{\boldmath $y$}) \; ,
\;\; \nabla^2_T \; H = 0 \, .
\label{LI23}
\end{equation}
Performing a  coordinate transformation,
we obtain the analog of (\ref{L10}):
\begin{equation}
ds^2 = ds_0^2 + 2 \k^2 E \theta (u)  du\, d\D_T \; ,
\label{LI24}
\end{equation}
and $(u, v, \mbox{\boldmath $y$})$ are asymptotically Cartesian
coordinates
provided $\mbox{\boldmath $\nabla$}H$ vanishes at large distances,
which requires that $H$
be asymptotically constant.  Being harmonic, it must then be a constant
everywhere,
and so merely
corresponds to  an irrelevant overall phase.
In contrast, for $D$=3 we have seen that no choice of homogeneous
solution will yield an
asymptotically
Cartesian metric; this clearly derives from the conical nature
of the exterior spatial geometry and the absence of the
corresponding Killing vectors even asymptotically \cite{012}.
[We can restate the above geometric discussion in terms of holonomy:
the photon's worldline divides the $u=0$ null hyperplane into the
$y<0$ and $y>0$ halfplanes.  The pure gravity solution is obtained
geometrically by making a cut along $u=0$ and then reidentifying
$u=0^-$ and $u=0^+$.  Points with $y<0$ ($y>0$) are reidentified
with a Lorentz transformation denoted $ {\cal L}_<$ (${\cal L}_>$).
The holonomy matrix associated with parallel transport around the source
is then given by ${\cal L} = {\cal L}_< {\cal L}_>^{-1} $, where
the first (second) factor is associated with crossing from (to)
$u<0$ to (from) $u>0$ on the $y>0$ ($y<0$) side.  The
coordinate transformation (\ref{L14b}) corresponds in the geometric picture
to the residual freedom associated with the holonomy-preserving transformation
$  {\cal L}_<  \rightarrow
 {\cal L}_<  \tilde{\cal L} $, ${\cal L}_> \rightarrow
{\cal L}_> \tilde{\cal L}$.]

A (pragmatic) resolution of this ambiguity follows if we are able
(on other grounds) to
find and justify appropriate boundary conditions to fix the
transverse
Green's function.  We have just seen that there is no compelling choice
within the framework of \cite{tH2}.  However
the boundary conditions {\it are} determined if we
go back
 to the derivation of these same results from the eikonal
approximation.
Indeed, as we will now show, the latter suggests that the
``correct'' transverse Green's function is to be determined from the
Feynman propagator for the exchanged
graviton.  Recall that the eikonal approximation in field theory
involves
summing the contributions of generalized ladder (of
exchanged gauge boson) diagrams to the $2\rightarrow 2$ particle
amplitude, keeping only the
leading ``hard momentum'' terms in the intermediate particle
propagators.
This is neatly done using functional techniques: take
two copies of the linearized $1 \rightarrow 1$ particle amplitude in an
external field (graphically just the sum of all numbers of graviton lines
emitted from a single particle line), and sew together the exchanged
gravitons in all possible ways using their Feynman propagator,
$(\Delta_F)_{\mu\alpha,\nu\beta}$.  For the
corresponding connected particle Green's functions this can be
summarized as
\begin{equation}
G_4(x_2',x_2;x_1',x_1) = \int{[dh_1]}[dh_2] G_2(x_2',x_2 | h_2)
G_2(x_1',x_1 | h_1) \exp \textstyle{\frac{1}{2}}
\int h_1^{\mu \nu}\Delta^{-1}_{\mu\alpha,\nu\beta}
h_2^{\alpha\beta} \, .
\label{LL1}
\end{equation}
Just the linearized theory is required \cite{KO14}, both in the
functional integral over $h$ and in calculating the particle Green's
functions;
the former since we sew with the free propagator, and the latter
because  --  consistent with keeping
only hard internal momenta  -- ladders are built with the 3-point
vertex alone.
[Strictly speaking, this would seem to conflict with (linearized) gauge
invariance,
but in our approximation, matter is essentially a fixed external
conserved source, so the problem is avoided.]
The problem is now to determine the 1-particle
propagator in a background linearized gravitational field, and thus the
remaining analysis reduces to that given for
QED in \cite{AI10,D11} (and our presentation will be essentially
equivalent to that in \cite{KO14}).
Following Schwinger (\cite{JS}, see also \cite{AI10,D11}) we introduce
formal  ``position and momentum observables'', $(X,P)$,
for the fast particle; then the single particle sector is
conveniently described by the Hamiltonian
$H(X,P)= H_0 + V$;
where  $H_0 = P^2$ and the potential is given (in
harmonic gauge) by $V = h_{\mu\nu}(x)P^{\mu}P^{\nu}$.
Thus $G_2= (H_0 + V)^{-1}$, and the 2-particle T-matrix is obtained as
the
on-shell limit of ${\cal T}_2 = H_0 (G_2 - H_0^{-1}) H_0$.
Proceeding under the approximation that the high energy particles'
momenta are essentially unchanged in the collision process, we may
replace
(working with particle 1 say)
$P \rightarrow p = (p_1 + p_1')/2$ in the resulting T-matrix.  The usual
exponentiation of the Born series then yields the eikonal approximation
\begin{equation}
<p_1'|{\cal T}^{(E)}|p_1> = \int\frac{d^3x}{(2\pi)^3}
e^{i(p_1 - p_1') \cdot x} \,
i\frac{d}{d\alpha} \, \exp \left[-\textstyle{\frac{i}{E}}
\int_{\alpha E}^{\infty}d\tau V(x+2\textstyle{\frac{p}{E}}\tau,p)
\right]
\big|_{\alpha = 0} \, .
\label{LL3}
\end{equation}
At this point it is useful to introduce new coordinates.
Consider the same kinematics as before: the fast particle
massless, with energy E.  We can write,
in the eikonal approximation, the null vector $p$ as
$p = E (1,\hat{\mbox{\boldmath $p$}})$,
and introduce the
vector $n = E (1, -\hat{\mbox{\boldmath $p$}})$ such that
$n\cdot p = -2E^2$.  Define
$ x^\mu = z^\mu + \textstyle{\frac{p^\mu}{E}} \sigma$,
where $n \cdot z = 0$.  In particular, then,
$(p_1 - p_1') \cdot x = q \cdot z$.  Moreover, $\sigma$ can be
shifted into the domain of the integral in the exponent, so that
the $\alpha$-derivative may be exchanged for a $\sigma$-derivative.
Putting this together, one easily finds that (\ref{LL3}) simplifies to
\begin{equation}
<p_1'|{\cal T}^{(E)}|p_1> =  2 i E \int\frac{d^2z}{(2\pi)^3} e^{iq \cdot
z}
e^{-2\textstyle{\frac{i}{E}}
\int_{-\infty}^{\infty}d\tau h_{\mu\nu}(z+2\textstyle{\frac{p}{E}}\tau)
p^\mu p^\nu}
\, .
\label{LL4}
\end{equation}
Having made the eikonal approximation for particle 1, we may simply
substitute
(\ref{LL4}) into (\ref{LL1}) (reduced on-shell), to obtain the result for
the $2 \rightarrow 2$ particle scattering amplitude.  In fact, the
$h$-integrals
apply a functional Taylor series expansion, and one finds
\begin{equation}
{\cal T}_4(p_1',p_1 ; p_2',p_2) =  2 i E \int\frac{d^2z}{(2\pi)^3}
e^{i q \cdot z}
{\cal T}_2(p_2',p_2 | h^C) \, ,
\label{LL5}
\end{equation}
where
\begin{equation}
h^C_{\mu\nu} =  \int{d^3}y (\Delta_F)_{\mu\alpha,\nu\beta}(x-y)
T^{\alpha\beta}(y)
\label{LL6}
\end{equation}
is the classical solution for the fast particle's energy-momentum source
$T_{\mu\nu}(y) = \frac{1}{2E} p_\mu p_\nu
\int{d\tau}\delta^3(y - z -  \frac{p}{E} \tau)$.
This is precisely of the expected form;
in particular the appropriate boundary conditions are clearly
fixed (through the choice of $\D_F$) {\it ab initio}.
[Note the additional integral
over particle 1's ``initial'' point, which simply incorporates
translational invariance and allows energy-momentum conserving
delta-functions to be factored out.]

Having ``justified'' the use of the Feynman propagator's boundary
conditions to define our $\D_T (y)$ of (\ref{L9}), we now evaluate the
resulting
scattering amplitudes, the simplest being that for Einstein gravity,
where the Feynman
propagator reduces to
the space-symmetric form
$\Delta_T^\sEG(y) = \textstyle{\frac{1}{2}} \kappa^2 |y|$.   The integral
 (\ref{L14a}) splits into two terms
which are easily calculated, and yield
\begin{equation}
T^\sEG(q_{y},s) = \frac{i}{(2\pi)^2}
\left(
\frac{1}{q_y + \kappa^2s/4} \: - \:
\frac{1}{q_y -\k^2s/4} \right) \sim
\frac{i}{2\pi^2} \:
\frac{\k^2s/4}{t+\k^4s^2/16}
\; .
\label{L16}
\end{equation}
In this theory, the amplitude
for the scattering of a test particle in the field due to another particle
has been calculated exactly \cite{tH212}, and at small
angles is just given by the result (\ref{L16}) (see also \cite{CIAF}
for a discussion of the full two-particle scattering amplitude).
One may well expect this since the geometry (with the conventional
sign of $\k^2$ used in (\ref{L16})) due to a particle in this
theory is flat with a conical singularity at the particle.  The
two terms in (\ref{L16}) correspond (roughly) to the test body's
passing on one side or the other of
the photon's cone.  For, note that in a suitably defined center of mass
system \cite{CIAF,KO36} $\sqrt{-t}$ is proportional to the scattering
angle,
$\th$, and $\kappa \sqrt{s}/2$ to the opening angle of the effective cone.
Thus
$\th_{cl} = \pm \kappa \sqrt{s}/2$ are the classical possibilities for the
angle
of scattering from this cone, and (\ref{L16}) just says that the
eikonal amplitude is dominated by $\th \sim \th_{cl}$.

It is interesting to pursue further the apparent relation of the above
result to
Aharonov--Bohm
(A-B) scattering.  In particular, let us consider the system which
certainly
does represent precisely that physical situation  -- point particles
coupled to an abelian vector Chern--Simons (CSE) action.  It is well
known
that the
constraints in this model imply that charged particles  also
carry a magnetic flux proportional to the charge,
and that this ``dressing'' is essentially
the only effect of the gauge field.  Thus, for example,  adiabatic
transport
of one particle around another picks up an A-B phase which
contributes to the statistics; these particles are anyons.
The two-particle scattering may be analyzed exactly of course, but we
may also
follow the procedure given above step by step.  In fact for later
convenience we will consider the vector system directly analogous to
TMG,
namely
topologically massive electrodynamics (TME) \cite{DJT6,S14}, for which
\begin{equation}
{\cal L} = - \textstyle{\frac{1}{4}}\, e^2 F_{\mu\nu}F^{\mu\nu} +
  \textstyle{\frac{1}{2}} \hth^{-1}
\epsilon^{\mu\nu\lambda}A_\mu\partial_\nu A_\lambda  + J^\mu
A_\mu \, ,
\label{L21}
\end{equation}
($F_{\mu\nu} = \partial_\mu A_\nu - \partial_\nu A_\mu$),
with field equations
\begin{equation}
e^{_2} \, \partial_\nu F^{\mu\nu} -
\hth^{-1} \epsilon^{\mu\nu\lambda}\partial_\nu A_\lambda =
J^\mu \, .
\label{L22}
\end{equation}
Note that $e^2$ has dimensions of mass in $D$=3, and $\hth$ is
dimensionless.  The  CSE theory is obtained in the limit
$e^2 \rightarrow \infty$, while Maxwell electrodynamics is the limit
$\hth \rightarrow \infty$.
As for gravity, we introduce the transverse Green's function
\begin{equation}
\left[ \frac{1}{e^2}\frac{d^2}{dy^2}  -
\hth^{-1} \frac{d}{dy} \right] \Delta^\sTME_T(y) = - \delta(y) \, ,
\label{L23}
\end{equation}
and the scattering amplitude is given by (\ref{L14a}), but without the
$s$  in the exponent (due to the lower spin
of the virtual exchanged boson).   The Feynman
propagator (in Landau gauge) is just
\begin{equation}
(\Delta_F^\sTME)_\mu^{\mu'} = \frac{e^2}{k^2 + \hth^{-2}e^4 -
i\epsilon}
\left[ P_\mu^{\mu'} + i \hth^{-1}e^2 \epsilon_{\mu}^{\,\,\a\m'}
\frac{k_\a}{k^2- i\epsilon}\right] + e^2 \frac{k_\m k^{\m'}}{k^4} \, ,
\;\;\;\; P^{\m '}_\m \equiv \d^{\m '}_\m - k_\m k^{\m '} k^{-2}.
\label{I4}
\end{equation}
Integrating against the fast particle current source we find,
in coordinate space, the particular solution
\begin{equation}
\Delta_T^\sTME(y) = \frac{\hth}{2} \epsilon(y) + \hth
e^{\hth^{-1}e^2y}
\th(-y) \, .
\label{I5}
\end{equation}
The CSE limit of (\ref{I5}) is smooth and the last term vanishes;
the calculation of the amplitude
is trivial, yielding the usual A--B result
\begin{equation}
T^\sCS(q_y) \sim \frac{1}{q_y} (e^{i \hat{\m}/2} - e^{-i \hat{\m}/2})
\, ,
\label{LLL1}
\end{equation}
where $q_y$ is the momentum transfer.
Note that $T^\sCS$ consists of two terms, which fail to cancel precisely
because of the A-B phase, and has the loose interpretation
that it sums the contributions from the particle's passing
``on one side or the other''  -- it is here that the analogy
goes through in (\ref{L16}).
The homogeneous solution ambiguities are quite different in
electrodynamics and in gravity, however.  For CSE, the
only
freedom is in a constant term in $\D_T$, which is of course irrelevant.
In pure Maxwell theory, the homogeneous term $Ay$ is also permitted,
and does lead to a different amplitude; however, because physics is
governed by the field strength, which (unlike the curvature) is a first
derivative, including this term actually corresponds to a different
physical
situation,
in which an additional source-free field is present to scatter the charge,
so that
there is no ambiguity in the physics here.

  There is another lesson to be learned from TME:
whereas the limit of the solution with Feynman propagator boundary
conditions to the (lower-derivative) CSE
is smooth, that in the
other, QED, direction encounters a singularity.  Expanding (\ref{I5}),
we see that
as $\hth \rightarrow \infty$, $\D_T$ tends to
$\hth/2 + e^2 y \th(-y)$.  This is the expected QED result
$-\frac{1}{2} e^2 |y|$, but only
up to the homogeneous solution
$H = \frac{1}{2} e^2 y + \hat{\m}/2$.
As is clear from (\ref{L14a}), the divergent constant term appears
in the amplitude as an irrelevant overall
phase.  However the linear term is curious given the ambiguity discussed above.
It is important to note that the behaviour of the amplitude in the limits is
strongly dependent on the infrared structure of the model.  For example, if
we introduce an infrared regulating mass for the vector field, we find
\begin{equation}
\Delta_T^M(y) = \frac{e^2}{\lambda} e^{-\hth^{-1}e^2 y/2}
e^{- \lambda |y|/2} \, , \,\,\, \lambda \equiv \sqrt{\hth^{-2}e^4 + 4 M^2} \, .
\label{i1}
\end{equation}
The infrared dependence is strikingly obvious from the fact that
limit $M \rightarrow 0$  does not reproduce (\ref{I5}).  However,
by keeping $ M \neq 0$ we find: $\hth \rightarrow \infty$ limits to the
correspondingly mass-regulated Green's function in QED, $e^2 e^{-M|y|}/(2M)$;
but for $e^2 \rightarrow \infty$ the result differs from the pure CSE one,
$\hth \epsilon(y)/2$, albeit only again by the ``infrared phase'' $\hth/2$.
Thus, although the model is, strictly speaking,
infrared regular -- as is to be expected since the excitation of TME is
massive -- this discussion shows that the infrared structure
may still be very subtle.  One obtains the flavour of what is going on directly
from the lowest order perturbative calculation of one vector exchange,
using the eikonal kinematics $p_1 \sim p_1'$, $p_1^u \sim p_2^v \sim 0$,
$q^u \sim q^v \sim 0$.  Then $p_1.p_2 \sim s$, while
$\epsilon^{\mu\alpha\nu} (p_1)_\mu q_\alpha (p_2)_\nu \sim q_y s$, and thus by
contracting (\ref{I4}) against the conserved
currents ({\it i.e.} $p_1^\mu (p_2)_{\mu'}$)
we find that in the dominant (t-channel) contribution the
odd-parity ``CS'' contribution is enhanced by the relative factor
of $q_y/(q_y^2 - i\epsilon)$ at small angles. However, for the
mass-regulated propagator precisely the opposite is true, since then
the factor becomes $q_y/(q_y^2 + M^2)$ and this term is suppressed at
small angles.  A little more work shows that such observations follow also
for the eikonal sum.  Using (\ref{I5}) and (\ref{L14a}),
and the following representation of the incomplete gamma function
\begin{equation}
\int dx \exp[- \alpha x - \beta e^{-x}] =
\beta^{-\alpha} \gamma(\alpha,\beta) \, ,
\label{i2}
\end{equation}
we find the TME amplitude
\begin{equation}
T^{\sTME} \sim \frac{e^{\hth/2}}{q_y} +
\frac{e^{-\hth/2}}{q_y} \beta^{-\alpha} \alpha \gamma(\alpha,\beta) \, ,
\label{i3}
\end{equation}
where $\alpha = i q_y \hth/e^2$ and $\beta = i \hth$.
The small angle (small $q_y$) expansion is then precisely the
large $e^2$ expansion in the second term (to make this quite obvious
the identity $\alpha \gamma(\alpha,\beta) = \gamma(\alpha + 1,\beta) +
\beta^\alpha e^{-\beta}$ is useful).

Returning to our scattering problem, the amplitude (\ref{L14a})
(with (\ref{L9})) of the
dynamical TMG theory has a similar complicated
expression in terms of the incomplete gamma function,
\begin{equation}
T^\sTMG(q_y) = \frac{i}{(2\pi)^2 } \frac{1}{q_y - \kappa^2 s/4}
e^{-\b} +
\frac{1}{(2\pi)^2 }\frac{1}{\mu}\: {(2\b)}^{-\a} \g (\a, 2 \b) e^\b \, ,
\label{L18}
\end{equation}
where
\begin{equation}
\a = \frac{i}{\mu} (q_y + \kappa^2 s/4) \, , \quad
\b = i \frac{\kappa^2}{4 \mu} s \, .
\label{L19}
\end{equation}
For large $\m$, the TMG amplitude can be expanded in powers of
$1/\m$
to provide Chern--Simons corrections to pure gravity, the leading one
being
\begin{equation}
 T^\sTMG(q_y,s) = \frac{i}{(2\pi)^2}
\left( \frac{\eta_+}{q_y+\k^2 s/4} -
\frac{\eta_-}{q_y-\k^2 s/4} \right) + {\cal O} (\m^{-2})
\label{L20}
\end{equation}
where $\eta_{\pm} = 1 \pm \b + {\cal{O}}({1\over \mu^2}).$
The effect of this first correction is to modify the otherwise equal
coefficients of the two A--B contributions in (\ref{L16}), as is to be
expected from its parity-violating character; note also that the expansion
brings in rising powers of $s$.

We can also obtain the amplitude in the CSG model, subject to the
caveat given earlier that the spacetime of (\ref{LI11}) is only fixed up to a
conformal factor;
if the scattered particle is also nearly null (in the same Lorentz frame),
this ambiguity becomes irrelevant.  We omit the details, but with the
choice (\ref{LI11})
and (\ref{L14a}), the amplitude may be written in terms of
the error function,
${\rm erf}(x) \equiv (2/\sqrt{\pi}) \int_0^x{dz} e^{-z^2}$, {\it i.e.}
\begin{equation}
T^\sCSG(q_y, s)  =
(2 \pi /s\tilde{\m})^{1/2}
[e^{i(\xi +\pi /4)}
\{1-{\rm erf} ((-i\xi )^{1/2})\} + {\rm c.c.}] \; , \;\;\;
\xi \equiv 4 q^2_y / (s\tilde{\m}) \; .
\label{i4}
\end{equation}
[It is possibly worth noting that to lowest order (\ref{i4}) can be put in
the form (\ref{L20}) but with different $\eta_\pm$, and with
$\kappa^2 s \rightarrow 2 \sqrt{\tilde{\mu} s}$,
although the significance of this remark is not clear.]

We are left with the question of limits of TMG solutions.  As in TME,
the limit to the lower derivative (here Einstein gravity) model is
smooth, as can be seen by considering the respective $\D_T$'s in the
two theories.  In fact, the full TMG Feynman propagator \cite{DJT6}
in  harmonic gauge reads
\begin{eqnarray}
(\Delta_F^\sTMG)_{\mu\nu}^{\mu'\nu'}(k) &=&
{\frac{\kappa^2}{2}}\frac{\mu^2}{k^2 + \mu^2 -i\epsilon}
\left[ - \left(\d_\mu^{(\mu'}\d_\nu^{\nu')} -
2 \eta_{\mu\nu}\eta^{\mu'\nu'}\right)\frac{1}{k^2 - i\epsilon}  -
\frac{1}{\mu^2}
\left(P_{\mu\nu}P^{\m'\n'} - 2 \eta_{\m\n}P^{\m'\n'}\right) \right.
\nonumber \\ [.15in]
& & + \left.
\frac{i}{2\mu} \frac{k_\a}{k^2}
\epsilon_{(\mu}^{\,\,\a(\m'}\d_{\n)}^{\n')}
+ \frac{1}{k^2 + 4\mu^2}\left(\frac{i}{2\mu}
\frac{k_{\a}}{k^2}k_{(\m}\epsilon_{\nu)}^{\,\,\a(\m'}k^{\n')} -
\frac{1}{k^2} k_{(\m}P_{\n)}^{\,\,(\m'}k^{\n')}\right)\right] \, ,
\label{I2}
\end{eqnarray}
where the round brackets denote (un-normalized) symmetrization.  This
clearly tends smoothly to the Einstein result (in the same gauge),
\begin{equation}
(\Delta_F^\sEG)_{\mu\nu}^{\mu'\nu'}(k) = -
{\frac{\kappa^2}{2}} \left(\d_\mu^{(\mu'}\d_\nu^{\nu')} -
2 \eta_{\mu\nu}\eta^{\mu'\nu'}\right)\frac{1}{k^2 - i\epsilon}
\label{I3}
\end{equation}
as $\mu \rightarrow \infty$.

The other limit---to CSG---of TMG solutions is, however, not smooth at
any level.
Indeed we have already seen that in this case the classical solution
only has a good limit if divergent homogeneous terms in the transverse
propagator are dropped.   The situation here is even worse than in
the TME case, where only the constant part of the homogeneous solution
diverged, and thus the divergence is absorbed in the usual exponentiated
phase.   Even then the limiting result differs from the QED result there
due to the remainder of the homogeneous solution (clearly a
drastic modification of the large $y$ behaviour), and it is {\it this}
which is changed by a regularization of the infrared behaviour.
The perturbative arguments which illuminated what was happening there
may also be applied for TMG, although in fact they do not
successfully carry over to the full eikonal sum.  Indeed for
TMG, as is evident from the full Feynman propagator, we may trace the
additional singularity to the fact that -- at tree
level -- CSG has an additional gauge invariance corresponding to Weyl
rescaling of the metric.  It is precisely the fate of this invariance when
coupled to interacting matter theories which makes the quantization of
CSG an interesting problem.  Typically we should expect no subtleties in
$D$=3.  One approach would be to maintain manifest
diffeomorphism invariance by obtaining the model as the limit of TMG, and yet
we have just seen some hints that this limit may be problematic.
This should reward further study.

\noindent{\bf Summary}

We have obtained the ``leading order'' two-particle gravitational
scattering amplitudes for three quite different gravity models in $D$=3
in the high $s$, small $t$ regime using the semi-classical
approximation
of \cite{tH2}, which is formally equivalent to that of the usual leading
order eikonal
expansion.  However, it is peculiar to $D$=3 that asymptotic flatness
is not sufficient to determine the Green's function completely.
  We have argued
heuristically in this respect that the eikonal
treatment gives a unique result, dictated by the use of the Feynman
propagator.  There is no such ambiguity in $D$=4, where {\it global}
asymptotically Cartesian coordinates
can always be introduced.  The resulting amplitude for $D$=3 Einstein
gravity was, as expected, of A--B form.  For full TMG, we also obtained
the scattering
amplitude but found no dramatic properties; it is not A--B
and indeed its corrections in an expansion about the Einstein value
``dephase'' the two characteristic A-B pieces of the latter.  We also
discussed the discontinuous aspects of the limiting process from TMG
to CSG at the level of solutions, and compared it with a similar
phenomenon in vector gauge theory.  At the classical level, the
impulsive plane wave
spacetimes generated by a null source that were obtained in TMG
provide the first explicit solution to full TMG due to a
localized source.

\vspace{.2in}

This work was supported by NSF grant PHY88--04561.
The research of AS was also supported by SERC at DAMTP,
while JM would like to acknowledge the support of the
Australian Research Council.

\end{document}